\documentclass[aps,prb,twocolumn,amsmath,amssymb,superscriptaddress,reprint,longbibliography, nofootinbib]{revtex4-2}

\usepackage[utf8]{inputenc}
\usepackage[T1]{fontenc}
\usepackage{graphicx}
\usepackage[colorlinks=true,citecolor=blue]{hyperref}
\usepackage{bbold}
\usepackage{xcolor}
\usepackage{threeparttable}
\usepackage{multirow}
\usepackage{natbib}
\usepackage[exponent-product = \cdot]{siunitx}
\usepackage{orcidlink}

\begin{document}
\renewcommand{\figureautorefname}{Fig.}
\renewcommand{\tableautorefname}{Tab.}
\renewcommand{\equationautorefname}{Eq.}
\title{Towards quantitative understanding of quantum dot ensemble capacitance-voltage spectroscopy}
\author{Nico F. Brosda\,\orcidlink{0009-0000-1479-5711}}
\thanks{These two authors contributed equally}
\author{Phil J. Badura\,\orcidlink{0009-0008-8567-224X}}
\thanks{These two authors contributed equally}
\author{İsmail Bölükbaşı} 
\author{İbrahim Engin}
\author{Patrick Lindner}
\author{Sascha R. Valentin\,\orcidlink{0000-0002-9078-1618}}
\author{Andreas D. Wieck\,\orcidlink{0000-0001-9776-2922}} 
\affiliation{Lehrstuhl für Angewandte Festkörperphysik, Ruhr-Universität Bochum, D-44801 Bochum, Germany}
\author{Björn Sothmann\,\orcidlink{0000-0001-9696-9446}}
\affiliation{Fakultät für Physik and CENIDE, Universität Duisburg-Essen, Lotharstraße 1, D-47048 Duisburg, Germany}
\author{Arne Ludwig\,\orcidlink{0000-0002-2871-7789}}
\email[Contact author: ]{arne.ludwig@rub.de}
\affiliation{Lehrstuhl für Angewandte Festkörperphysik, Ruhr-Universität Bochum, D-44801 Bochum, Germany}
\date{\today}

\begin{abstract}

Inhomogeneous ensembles of quantum dots (QDs) coupled to a charge reservoir are widely studied by using, e.g., electrical methods like capacitance-voltage spectroscopy. We present experimental measurements of the QD capacitance as a function of varying parameters such as ac frequency and bath temperature. The experiment reveals distinct shifts in the position of the capacitance peaks. While temperature-induced shifts have been explained by previous models, the observation of frequency-dependent shifts has not been explained so far.
Given that existing models fall short in explaining these phenomena, we propose a refined theoretical model based on a master equation approach which incorporates energy-dependent tunneling effects. This approach successfully reproduces the experimental data. We highlight the critical role of energy-dependent tunneling in two distinct regimes: at low temperatures, ensemble effects arising from energy-level dispersion in differently sized QDs dominate the spectral response; at high temperatures and frequencies, we observe a peak shift of a different nature, which is best described by optimizing the conjoint probability of successive in- and out-tunneling events.
Our findings contribute to a deeper understanding of tunnel processes and the physical properties of QD ensembles coupled to a common reservoir, with implications for their development in applications such as single-photon sources and spin qubits.

\end{abstract}

\maketitle


\section{Introduction}

Quantum information science, encompassing applications such as quantum communication, cryptography, and computation, is progressing rapidly towards practical realization \cite{senellartHighperformanceSemiconductorQuantumdot2017a, lodahlQuantumdotBasedPhotonic2017, huberSemiconductorQuantumDots2018, uppuScalableIntegratedSinglephoton2020}.
Among the various quantum hardware platforms, semiconductor quantum dots (QDs) stand out due to their ability to confine single charge carriers with atom-like discrete energy levels, making them highly suitable for optoelectronic applications \cite{najerGatedQuantumDot2019, liuSolidstateSourceStrongly2019, keilSolidstateEnsembleHighly2017}.
In particular, InAs QDs exhibit fully quantized energy states similar to atoms \cite{drexlerSpectroscopyQuantumLevels1994, gammonFineStructureSplitting1996} and can be coupled to multiple solid-state degrees of freedom, such as their dielectric environment \cite{franceschettiAdditionSpectraQuantum2000, kerskiQuantumSensorNanoscale2021}, photonic modes \cite{reithmaierStrongCouplingSingle2004, gangloffQuantumInterfaceElectron2019, zhaiQuantumInterferenceIdentical2022}, phononic interactions \cite{besombesAcousticPhononBroadening2001, hohenesterPhononassistedTransitionsQuantum2009}, nuclear spin systems \cite{stockillQuantumDotSpin2016, jacksonOptimalPurificationSpin2022, chekhovichNuclearSpinEffects2013}, and electronic states or charge reservoirs \cite{warburtonCoulombInteractionsSmall1998, kroutvarOpticallyProgrammableElectron2004}.
These couplings present both challenges and opportunities \cite{gillardFundamentalLimitsElectron2021}.

Capacitance-voltage (C-V) spectroscopy is a well-established technique for probing the energy level structure of semiconductor QDs \cite{ashooriSingleelectronCapacitanceSpectroscopy1992, ashooriNelectronGroundState1993}. 
By applying a gate voltage with a superimposed ac modulation, charge fluctuations in a typically diode-like structure are induced and detected, enabling the investigation of discrete charging events and quantum confinement effects \cite{drexlerSpectroscopyQuantumLevels1994}.
The dynamic nature of C-V spectroscopy offers valuable insights into tunneling processes \cite{luykenDynamicsTunnelingSelfassembled1999} and elucidates, e.g., how the state degeneracy of QD levels affects tunneling \cite{beckelAsymmetryChargeRelaxation2014}.
This is particularly relevant for understanding the observed tunnel resonance \cite{brinksThermalShiftResonance2016}, where a thermal shift in the quasi-equilibrium measurements of the lowest energy \textit{s}-states has been described.
Additional insights have been obtained using a master equation approach \cite{beenakkerTheoryCoulombblockadeOscillations1991}, which extends to non-equilibrium states \cite{valentinIlluminationinducedNonequilibriumCharge2018}.

The master equation approach of Ref. \cite{valentinIlluminationinducedNonequilibriumCharge2018} assumed all QDs to have identical properties. It considered a QD system that was either empty or occupied by an electron or hole, with tunnel rates describing the transitions between levels.
The dynamics described by this master equation helped to understand C-V experiments and provided insights into various phenomena, such as the occurrence of illumination-induced nonequilibrium peaks.

In this paper, we present experimentally obtained C-V spectra that cannot be fully explained by the previously described model and, thus, require an extension of the theoretical framework. In particular, we demonstrate that a refined model has to consider at least two additional features. First, one has to account for the energy-dependent nature of tunnel coupling, arising from the variation of the barrier length between the QD and the reservoir with energy tuning. Second, the dispersion in QD sizes affects their energy levels, potentially leading to different behavior in QD ensembles compared to individual QDs.

The paper is organized as follows. In Sec.~\ref{sec:exp}, we present our experimental results for the C-V spectra of a QD ensemble for different temperatures and ac frequencies. In Sec.~\ref{sec:theory}, we develop our master-equation approach including energy-dependent tunnel couplings. Within our theoretical approach, we calculate C-V spectra that are subsequently compared to our experimental results. Conclusions to advance towards a more comprehensive understanding of tunneling dynamics, specifically C-V spectroscopy in QD ensembles, are drawn in Sec.~\ref{sec:conclusion}.


\section{\label{sec:exp}Experiment and Results}

The foundation for the subsequent theoretical analysis is established through experimental observations of a QD ensemble coupled to a charge reservoir. Capacitance-voltage spectroscopy is employed to investigate QDs embedded in a GaAs semiconductor matrix. Initially, we will discuss the fundamental aspects of the sample structure and C-V spectroscopy, as these details define the context for the later theoretical discussions.

\begin{figure}
    \centering
    \includegraphics[width=\columnwidth]{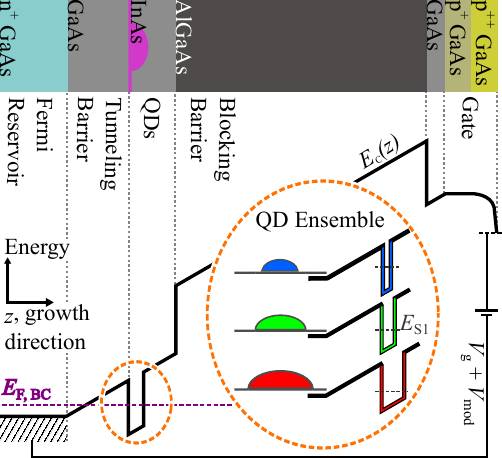}
    \caption{Schematic representation of the sample layers and the corresponding 1D conduction band edge. Electrons from the doped Fermi reservoir can tunnel into the QD as soon as the gate voltage $V_\mathrm{g}$ aligns the QD energy levels. The tunneling barrier in between reservoir and QD is assumed to have a triangle shape. A blocking barrier prevents charge drift towards the top gate on top. Additionally, the sketch illustrates a QD ensemble depicted by potential wells of varying sizes, reflecting the different QD sizes observed in the experiment. Note that the sample dimensions are not shown to scale.}
    \label{fig:sketch_bandstructure}
\end{figure}

\autoref{fig:sketch_bandstructure} shows the sample structure and a sketch of the corresponding conduction band edge.
The analyzed sample is grown via molecular-beam epitaxy\footnote{Same sample as in reference \cite{valentinIlluminationinducedNonequilibriumCharge2018}; for internal use and identification: Sample {\#}14729c}.
QDs are embedded in a GaAs n-i-p diode structure. A Si-doped GaAs layer (n-doping density $N_{D} = \text{\SI{2e+18}{\per\cubic\centi\meter}}$) acts as the charge reservoir.
Carriers need to surpass the GaAs layer (\SI{35}{\nano\meter}) following on top of the charge reservoir to access the energetic states within a QD.
Therefore, this tunnel barrier plays a crucial role in the tunnel-coupling between reservoir and QD.
QDs are formed by Stranski-Krastanov growth of InAs on the GaAs matrix and are capped by intrinsic GaAs (\SI{11}{\nano\meter}).
Above the QDs a blocking barrier composed out of 50 periods of Al$_{0.33}$Ga$_{0.67}$As/GaAs (\SI{3}{\nano\meter}/\SI{1}{\nano\meter}, for simplicity depicted with uniform material in \autoref{fig:sketch_bandstructure}) prevents carrier diffusion towards the top gate, which serves as the second electric contact.

The epitaxial top gate is formed out of bulk carbon-doped GaAs (p-doping density $N_{A} = \text{\SI{3e+18}{\per\cubic\centi\meter}}$; \SI{25}{\nano\meter}) and 40 periods of carbon-delta-doped and carbon-doped GaAs ($N_{A} = \text{\SI{1e+19}{\per\cubic\centi\meter}}$; total thickness of \SI{20}{\nano\meter}).
Due to the high doping we form an Ohmic contact by simply bonding to the top gate.
With standard wet chemical etching mesas of $\text{300} \times \text{\SI{300}{\micro\square\meter}}$ are processed.
The back-contact to the reservoir is formed by soldering indium to the corners of a $\text{4} \times \text{\SI{5}{\milli\square\meter}}$ sample.

QD formation by Stranski-Krastanov growth of InAs is a probabilistic process, resulting in QDs of varying sizes.
Consequently, the sample contains an ensemble of QDs with dispersed energy levels.
For the sample analyzed in this paper, the ensemble width is around \SI{100}{\milli\electronvolt}.
In C-V spectroscopy, we examine the behavior of this ensemble, as the top gate covers multiple QDs (approximately a QD density of \SI{5e+9}{\per\square\centi\meter}; meaning \num{4.5e+6} QDs under one mesa).

Applying a voltage to the top gate shifts the energy levels, particularly those inside the QDs, relative to the Fermi level defined by the charge reservoir.
The gate voltage can thus tune different quantized states of the QDs in resonance with the Fermi energy of the reservoir, enabling the energy-spectroscopic nature of C-V measurements.
However, this static shift alone does not permit direct detection of these energy levels.

\begin{figure}
    \centering
    \includegraphics[width=\columnwidth]{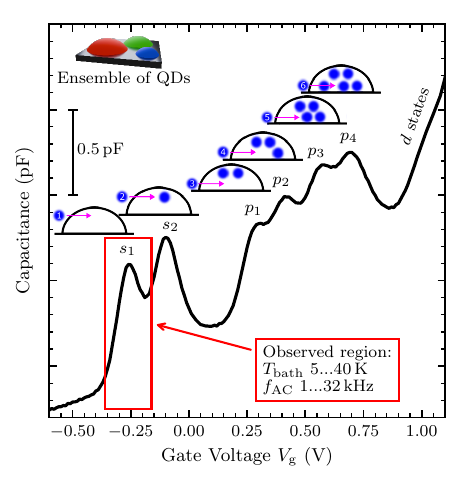}
    \caption{Exemplary C-V spectrum obtained from the sample with structure corresponding to \autoref{fig:sketch_bandstructure}. The measured capacitance is plotted as a function of the gate voltage $V_\mathrm{g}$. The peaks in this spectrum correspond to resonant coupling with energy levels inside the QD. For reference, the levels and their associated peaks are denoted as \textit{s}1 for the first electron occupying the lowest level, p1 for the third electron occupying the second level, and so on. In this study, we focus specifically on the 
    \textit{s}1 peak, highlighted by a red box. C-V spectra of this type were recorded at various bath temperatures and different frequencies for the ac component of the gate voltage $U_\mathrm{mod}$.}
    \label{fig:example_C-V}
\end{figure}

To detect different charge states and energy levels in a QD, a dynamic ac voltage added to the static component of the gates voltage is mandatory.
If an energy level of the QDs is aligned with the Fermi level of the charge reservoir of the sample, tunnel processes lead to a charging of the QDs which result in capacitance peaks.
Such a C-V spectrum from a sample with a structure as introduced above is shown in \autoref{fig:example_C-V}.

The C-V spectrum in \autoref{fig:example_C-V} displays several peaks superimposed on an increasing background.
This background arises from the diode-like nature of the sample structure and will be subtracted when precise peak positions are evaluated.
Peaks emerge from this background whenever the QD energy levels come in resonance with the Fermi level of the charge reservoir.
Charging the QD is a sequential process where Coulomb repulsion lifts the degeneracy of energy levels.
Two distinct peaks are observed for the twofold-degenerate lowest energy level of the QD, labeled \textit{s}1 (first electron) and \textit{s}2 (second electron).
The next higher energy level shows four peaks, \textit{p}1, \textit{p}2, \textit{p}3, and \textit{p}4, corresponding to the third, fourth, fifth, and sixth electron charged in the QD, respectively.
Peaks associated with even higher \textit{d}-states are barely discernible from the background for the highest gate voltages shown in \autoref{fig:example_C-V}.

The following analysis will focus primarily on a single peak in the C-V spectra and its shift to different gate voltages.
Thus, it is essential to understand what the maximum of a C-V peak signifies.
Whenever an energy state comes into resonance with the Fermi level of the charge reservoir, the driven charge exchange between the QD and the reservoir gives rise to a maximum of the capacitance.
Another important consideration involves the ensemble of QDs measured in the example shown in \autoref{fig:example_C-V}.
A peak here does not represent one tunnel process into a single QD but rather tunneling into many QDs of varying sizes.
This not only broadens the C-V peak but may also affect the peak position.

The experimental findings in this work will focus on the \textit{s}1 peak (as marked in \autoref{fig:example_C-V}), which corresponds to the first electron tunneling in and out of the QD. C-V spectra in this region were recorded using a helium-based closed-cycle cryostat, which allows for temperature control ranging from \SIrange{2.5}{300}{\kelvin}. For our measurements, we set the temperature and perform C-V measurements at various ac frequencies. The signal current through the back contact is detected using a lock-in amplifier.

\begin{figure}
    \centering
    \includegraphics[width=\columnwidth]{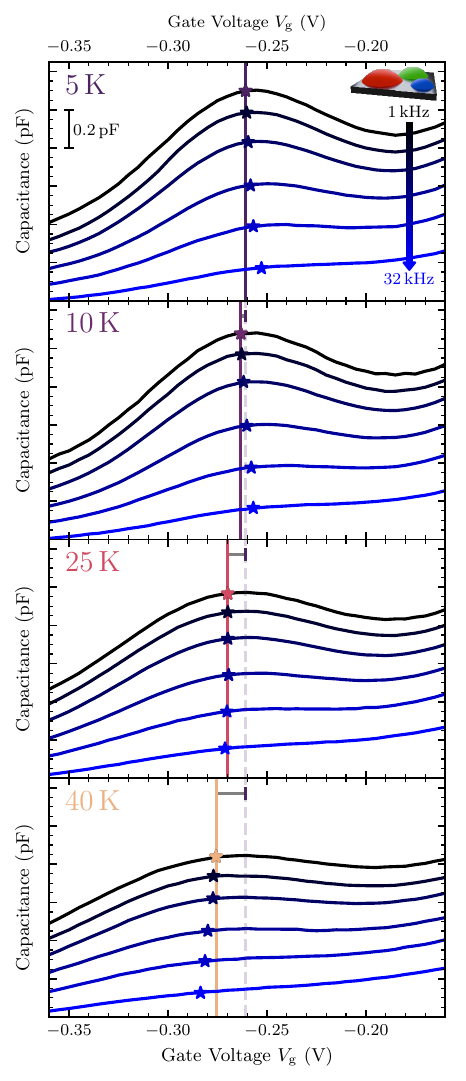}
    \caption{C-V spectra of the QDs' \textit{s}1 state at different temperatures from \SIrange{5}{40}{\kelvin}. Each spectra is measured from low (black) to high (blue) ac frequencies ($\approx\,$\SIrange{0}{34}{\kilo\hertz}). Vertical lines mark the peak position at the lowest measured frequency. At each frequency, the peak was fitted and its maximum is marked by a star. This highlights the shift induced by temperature and frequency. Note that the curves have been aligned to a distance of \SI{0.07}{\pico\farad} in between their maxima for better visibility.}
    \label{fig:experimental_findings}
\end{figure}

\autoref{fig:experimental_findings} shows the capacitance as a function of gate voltage for different frequencies at several temperatures ranging from \SIrange{5}{40}{\kelvin}. The positions of the peaks, marked for various frequencies from \SIrange{1}{32}{\kilo\hertz}, are determined by fitting the \textit{s}1 and \textit{s}2 peaks with two Gaussian functions (as we model the ensemble by a Gaussian distribution) in the spectra after subtracting the underlying diode characteristic. Vertical lines indicate the positions at the lowest measured frequency. The capacitance scale is consistent across all graphs. Note that the curves have been intentionally shifted for better comparison.

We first investigate the lowest frequency of \SI{1}{\kilo\hertz} between the different panels represented by the upper black curve. As the temperature increases, we observe a broadening of the \textit{s}1 peak, which is common and expected. Additionally, we see a thermal shift in the peak position towards lower gate voltages (corresponding to higher energies). This shifting behavior occurs as the peak position is determined by an equality of the rates for tunneling in and out of the QD and was already observed and discussed \cite{brinksThermalShiftResonance2016}.

We now turn to the higher frequencies. As the frequency increases, the peak height decreases. This suppression occurs because fewer electrons are able to tunnel into the QDs during one period of the applied ac frequency. Consequently, the charge current diminishes, leading to a reduction in the measured capacitance. In addition to the suppression of peak height, the peak position is also affected by rising frequency. At the lowest temperature of \SI{5}{\kelvin}, the \textit{s}1 peak shifts toward higher gate voltages (corresponding to lower energies). 

Surprisingly, this frequency-induced shift completely changes its behavior at higher temperatures. At an intermediate temperature of \SI{25}{\kelvin}, the frequency-dependent shift appears to vanish. Furthermore, for the highest temperature of \SI{40}{\kelvin} the \textit{s}1 peak shifts into the opposite direction, towards lower voltages. 

Such a shifting behavior has not been observed or explained before. The discovery of this effect suggests that the C-V spectroscopy of charge-reservoir-coupled QDs, and therefore the characteristics of the tunneling processes between the reservoir and the QDs, are not yet fully understood. To move towards a quantitative understanding of QD ensemble capacitance-voltage spectroscopy, we will now proceed with developing a model that accounts for these experimental observations.


\section{\label{sec:theory}Theory and model}

\subsection{Mathematical description of C-V spectroscopy}
In the following, we develop a model based on a master-equation approach which allows us to explain our experimental results presented above. We assume that a time-dependent gate voltage
\begin{equation} \label{eq:theory:signal}
    V(t) = V_\mathrm{g} + \varepsilon \cos(\omega t)
\end{equation}
is applied to the sample where $V_g$ is the static gate voltage, $\varepsilon$ denotes the ac amplitude and $\omega$ is the angular frequency with the corresponding period $T=2\pi/\omega$. The time-dependent gate voltage gives rise to a periodic modulation of the level positions in the QD which in turn leads to a periodic tunneling in and out of electrons via the back contact. As a result, the charge $Q(t)$ on the QD fluctuates and gives rise to an ac current $I(t)=-\dot Q(t)$. In the following, we are mainly interested in the first harmonic of this current which is given by
\begin{equation} \label{eq:theory:fourier}
    I_1 = \frac{2}{T} \!\int_0^T \!\!\mathrm{d}t \,\mathrm{e}^{-\mathrm{i} \omega t} I(t).
\end{equation}

The charge $\mathrm{d}Q$ tunneling into the QD from the back contact is related to the change $\mathrm{d}V$ in the gate voltage by the differential capacitance $C(V)$ via
\begin{equation} \label{eq:theory:differential_capacitance}
    -\mathrm{d}Q = C(V)\,\mathrm{d}V .
\end{equation}
We can therefore express the ac current as
\begin{equation} \label{eq:theory:current_relation}
    I = -C(V) \,\varepsilon \omega \sin(\omega t),
\end{equation}
such that to first-order in $\varepsilon$ we have
\begin{equation} \label{eq:theory:connect_experiment_theory}
    C(V_\mathrm{g}) = \tfrac{\mathrm{i}}{\omega\varepsilon} I_1 + \mathcal{O}(\varepsilon^2).
\end{equation}
\autoref{eq:theory:connect_experiment_theory} allows us to connect the differential capacitance $C(V_g)$ to the experimentally measurable quantity $I_1$ which can also be evaluated theoretically from the time-dependent QD charge $Q(t)$.

\subsection{Master-equation approach}
In order to model the charging dynamics of a single QD, we extend the master equation model presented by \citeauthor{valentinIlluminationinducedNonequilibriumCharge2018} \cite{valentinIlluminationinducedNonequilibriumCharge2018}.
Every state $s$ of a QD has a time-dependent occupation probability $p_s(t)$ accompanied by a corresponding net charge $q_s$.
Introducing the vector quantities $\boldsymbol{p}(t) = ( p_s(t) )$ and $\boldsymbol{q} = ( q_s )$ enables us to write the QD charge as
\begin{equation} \label{eq:theory:quantum_dot_charge}
    Q(t) = \boldsymbol{q}^T \boldsymbol{p}(t) = \textstyle\sum_s q_s p_s(t).
\end{equation}
The time-dependent probability vector $\boldsymbol{p}(t)$ obeys the master equation
\begin{equation} \label{eq:theory:master_equation}
    \dot{\boldsymbol{p}}(t) = \boldsymbol{W} \boldsymbol{p}(t) ,
\end{equation}
where $\boldsymbol{W}$ denotes the matrix of tunneling rates. Its off-diagonal elements are given by the transition rates from state $s$ to $s'$ while the diagonal elements are determined by the conservation of probability
\begin{equation} \label{eq:theory:conservation_of_probability}
    \boldsymbol{1}^T \boldsymbol{W} = 0 \quad\text{with}\quad \boldsymbol{1} = (1, \dots, 1)^T .
\end{equation}
We assume our problem \autoref{eq:theory:master_equation} to be sufficiently regular to have a unique periodic solution representing a steady state.
Further assuming analytic dependence on small $\varepsilon$, we expand $\boldsymbol{W}$ in a Taylor series around $\varepsilon = 0$, i.e.,
\begin{equation}
    \boldsymbol{W} = \sum_{n = 0}^\infty \boldsymbol{W}^{(n)} \varepsilon^n \cos(\omega t)^n, \quad \boldsymbol{W}^{(n)} = \frac{1}{n!} \frac{\partial^n \boldsymbol{W}}{\partial V_\mathrm{g}^n} ,
\end{equation}
and expand $\boldsymbol{p}$ itself as the power series
\begin{equation} \label{eq:theory:p_ansatz}
    \boldsymbol{p}(t) = \sum_{n = 0}^\infty \boldsymbol{p}^{(n)}(t) \,\varepsilon^n .
\end{equation}
Substituting into the master equation \autoref{eq:theory:master_equation} and collecting terms of order $n$ yields the inhomogeneous equations
\begin{equation} \label{eq:theory:master_equation_ansatz_order}
    (\partial_t - \boldsymbol{W}^{(0)})\, \boldsymbol{p}^{(n)}(t) = \boldsymbol{r}^{(n)}(t)
\end{equation}
with the inhomogeneities
\begin{equation} \label{eq:theory:residues}
   \boldsymbol{r}^{(n)}(t) = \sum_{l = 1}^n \boldsymbol{W}^{(l)} \boldsymbol{p}^{(n - l)}(t) \cos(\omega t)^l .
\end{equation}
As we are only interested in the periodic stationary case, we
expand
\begin{align}
	\boldsymbol{p}^{(n)}(t)=\sum_k \boldsymbol{p}^{(n)}_k e^{ik\omega t},\ \ \boldsymbol{r}^{(n)}(t)=\sum_k \boldsymbol{r}^{(n)}_k e^{ik\omega t},
\end{align}
such that we can express \autoref{eq:theory:master_equation_ansatz_order} as
\begin{equation} \label{eq:theory:master_equation_ansatz_order_fourier}
    (\mathrm{i} \omega k - \boldsymbol{W}^{(0)})\, \boldsymbol{p}^{(n)}_k = \boldsymbol{r}^{(n)}_k .
\end{equation}
When $k \neq 0$, \autoref{eq:theory:master_equation_ansatz_order_fourier} is solved by inverting $(\mathrm{i} \omega k - \boldsymbol{W}^{(0)})$, which is always possible due to Gershgorin's circle theorem.
The case $k = 0$ poses a singular equation, as \autoref{eq:theory:conservation_of_probability} prevents $\boldsymbol{W}^{(0)}$ to be invertible.
This singularity requires additional constraints, which come in the form of the normalization of probability $\boldsymbol{1}^T \boldsymbol{p}(t) = 1$.
Thus, for $n = 0$, we obtain the eigenvalue problem
\begin{equation} \label{eq:theory:master_equation_0}
    \boldsymbol{W}^{(0)} \boldsymbol{p}^{(0)}_k = \boldsymbol{0}, \quad\boldsymbol{1}^T \boldsymbol{p}^{(0)}_k = \delta_{k0} .
\end{equation}
Otherwise, these normalization constraints demand $\boldsymbol{1}^T \boldsymbol{p}^{(n)}_0 = 0$, such that adding $\boldsymbol{p}^{(0)} \boldsymbol{1}^T \boldsymbol{p}^{(n)}_0 = 0$ to \autoref{eq:theory:master_equation_ansatz_order_fourier} leaves it invariant under the given constraints.
Combining the preceding observations allows us to write
\begin{equation} \label{eq:theory:master_equation_sol_0}
    \boldsymbol{p}^{(n)}_k = (\mathrm{i} \omega k - \boldsymbol{W}^{(0)} + \delta_{k 0} \,\boldsymbol{p}^{(0)} \boldsymbol{1}^T)^{-1}\,\boldsymbol{r}^{(n)}_k ,
\end{equation}
as $\boldsymbol{p}^{(0)} \boldsymbol{1}^T - \boldsymbol{W}^{(0)}$, unlike $\boldsymbol{W}^{(0)}$, is actually invertible assuming a one-dimensional nullspace.
Using \autoref{eq:theory:master_equation_sol_0} one can successively compute higher orders of $\boldsymbol{p}$ while collecting Fourier modes in \autoref{eq:theory:residues}.
With \autoref{eq:theory:connect_experiment_theory} in mind, we are mostly interested in the case $k = 1$.
Performing these calculations up to order $n = 2$, and inserting them into \autoref{eq:theory:quantum_dot_charge} and \autoref{eq:theory:connect_experiment_theory}, leads to
\begin{equation} \label{eq:theory:capacity_from_master_equation}
    C(V_\mathrm{g}) = -\boldsymbol{q}^T (\mathrm{i} \omega - \boldsymbol{W}^{(0)})^{-1}\,\boldsymbol{W}^{(1)} \boldsymbol{p}^{(0)} + \mathcal{O}(\varepsilon^2) .
\end{equation}
As our experimental data corresponds to the real part of $C(V_\mathrm{g})$ we will only consider the real part of \autoref{eq:theory:capacity_from_master_equation}.

\subsection{Tunneling into the \textit{s}1 state}
We proceed by modelling the \textit{s}1 peak by the simple two state model of a QD being filled by one or no electron.
Denoting the rate for tunneling into the QD as $\Gamma_\mathrm{in}$ and tunneling out as $\Gamma_\mathrm{out}$, we accordingly consider
\begin{equation}
    \boldsymbol{W} =
    \left(
    \begin{array}{cc}
        -\Gamma_\mathrm{in} & +\Gamma_\mathrm{out} \\
        +\Gamma_\mathrm{in} & -\Gamma_\mathrm{out}
    \end{array}
    \right)
    , \quad
    \boldsymbol{q} =
    \left(
    \begin{array}{r}
        0 \\
        \!-e
    \end{array}
    \right) ,
\end{equation}
where $e$ is the elementary charge.
As previously described by Ref. \cite{brinksThermalShiftResonance2016}, the tunneling rates are given by
\begin{align}
    \Gamma_\mathrm{in} &= g_\mathrm{in} \,\mathcal{T}(E)\,f(E)\,D(E) \, \label{eq:theory:tunneling_rate_in} \\
    \Gamma_\mathrm{out} &= g_\mathrm{out} \,\mathcal{T}(E) \,[1 - f(E)]\,D(E) , \label{eq:theory:tunneling_rate_out}
\end{align}
where $f(E)$ symbolizes the Fermi-Dirac distribution
\begin{equation} \label{eq:theory:fermi_dirac}
    f(E) = \frac{1}{\mathrm{e}^{E / k_\mathrm{B} T} + 1}, \quad E = E_{s1} - E_\mathrm{F},
\end{equation} 
with the energy $E_{s1}$ of the \textit{s}1 state and the Fermi level $E_\mathrm{F}$.
$D(E)$ denotes the density of states and $f(E) D(E)$ quantifies the probability of an electron being available for tunneling into the QD; $[1 - f(E)]D(E)$ holds the respective complementary probability of finding an unoccupied state for tunneling back.
The degeneracies for these processes are given by $g_\mathrm{in}$ and $g_\mathrm{out}$, and the energy-dependent tunnel coupling strength by $\mathcal{T}(E)$.

Before discussing the broader implications of these expressions, we will first specify the particular tunneling rates used to illustrate the general observations and to fit the experimental data.

\subsection{Derivation of the tunnel coupling}

When the first electron tunnels into the QD, its spin orientation is arbitrary, leading to a degeneracy factor $g_\mathrm{in} = 2$ as both spin-up and -down electrons contribute to tunneling.
Conversely, an electron in the QD has a definite spin, resulting in $g_\mathrm{out} = 1$.

\begin{figure}
    \centering
    \includegraphics[width=\columnwidth]{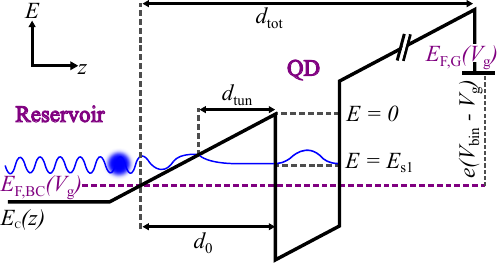}
    \caption{Sketch of the conduction band edge between charge reservoir and QD. $d_0$ names the distance between reservoir and QD, $d_\mathrm{tot}$ the length between reservoir and top gate. The effective tunnel length $d_\mathrm{tun}$ can be calculated by an approach corresponding to the WKB-approximation.}
    \label{fig:WKB}
\end{figure}

To model the tunnel coupling strength $\mathcal{T}(E)$, we assume that the voltage-dependent deformation of the conduction band edge $E_\mathrm{C}$ can be effectively described using a simple lever approach \cite{warburtonCoulombInteractionsSmall1998, leiProbingBandStructure2008} as shown in \autoref{fig:WKB}.
Note that for convenience in the following calculations the band edge at the QD is taken as constant, making the Fermi levels at gate $E_\mathrm{F,G}$ and back contact $E_\mathrm{F,BC}$ functions of the applied gate voltage $V_\mathrm{g}$.
In this convention, a varying gate voltage $V_\mathrm{g}$ modifies the relative position of the Fermi level in the QD.
Denoting the lever by $\lambda$, we find
\begin{equation} \label{eq:theory:lever}
    E_\mathrm{F}(V_\mathrm{g}) := E_\mathrm{F, BC}(V_\mathrm{g}) = \frac{e}{\lambda} (V_\mathrm{g} - V_\mathrm{bin}) ,\ \ \lambda = \frac{d_\mathrm{tot}}{d_0} 
\end{equation}
with $d_\mathrm{tot}$ and $d_0$ being the distance between back contact and top gate or QD, respectively.
Next, we approximate $\mathcal{T}(E)$ by estimating the probability density $|\Psi|^2$ of the reservoir electrons at the left QD boundary using a one-dimensional WKB approximation for the wavefunction $\Psi$.
Using basic geometry, one can find the triangular well potential $U(z)$ and the effective tunnel distance $d_\mathrm{tun}$ between reservoir and QD, hence
\begin{equation} \label{eq:WKB}
    \mathcal{T} \!\propto\! |\Psi|^2 \!\propto \mathrm{e}^{-2 {\textstyle\int_0^{d_\mathrm{tun}}} \frac{\sqrt{2 m^\ast (U \text{-} E_{s1})}}{\hbar} \mathrm{d}z}\! = \mathrm{e}^{\frac{4 d_0 \sqrt{2 m^\ast}}{3 \hbar} \frac{\sqrt{\text{-}E_{s1}}^3}{E_\mathrm{F}}} \!\!,\!
\end{equation}
where $m^\ast$ refers to the effective mass of GaAs and $E_{s1}$ to the energy of the QD's lowest quantized state.
The density of states $D(E)$ can be assumed constant, as tunneling occurs from conduction band edge electronic states.

\begin{figure}
    \centering
    \includegraphics[width=\columnwidth]{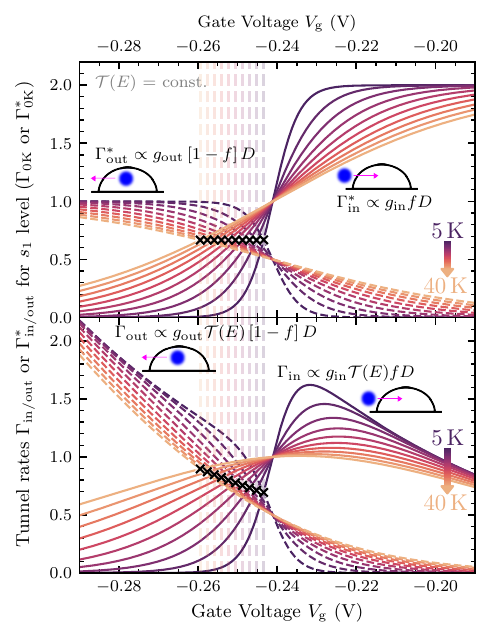}
    \caption{Tunneling rates into (solid) and out of (dashed) a QD's \textit{s}1 state ($V_\mathrm{g} = \SI{-0.23}{\volt}$) for various temperatures and for considering no energy dependence of the tunneling barrier ($\mathcal{T}(E) = \mathrm{const.}$) or the in \autoref{eq:WKB} derived dependence for $\mathcal{T}(E)$. Note that in the energy independent case the tunneling rates are given by a scaled Fermi distribution. The tunnel-in rates remain zero, until the quantized energy state comes in resonance with the reservoir. This onset becomes smeared at higher temperatures due to its dependence on the Fermi distribution. Tunneling out of the QDs occurs at a high rate under negative bias, sharply dropping when the QD energy level aligns with the reservoir's Fermi energy - Pauli principle inhibits tunneling into occupied states. Again, this is washed out at elevated temperatures. The intersections $\Gamma_\mathrm{in}=\Gamma_\mathrm{out}$ are marked with a cross.}
    \label{fig:tunnel_rates}
\end{figure}

Using these information we are able to look at the tunneling rates for in and out tunneling in reference to the QD's \textit{s}1 energy state. \autoref{fig:tunnel_rates} shows these rates calculated for the experimentally observed sample structure and in both cases, considering energy dependence in tunneling or not. All rates are normalized to the static result for the rate at \SI{0}{K} and are shown in dependence of the gate voltage and for different temperatures. If we neglect the energy dependence, the rates are simply given by a Fermi distribution, scaled by the degeneracies for tunneling in and out. If we consider the dependence on $\mathcal{T}(E)$ found in \autoref{eq:WKB}, this leads to a modification of these curves. The tunneling-out rate shows a steep drop over the whole voltage range, while the tunneling-in rate now decreases after a maximum, with increasing gate voltage.

In both cases depicted in \autoref{fig:tunnel_rates}, the temperature dependence is governed by the Fermi distribution in the tunnel rates. 
For the observed low-temperature regime, the temperature dependence of the chemical potential can be neglected such that the point where $f(E) = 0.5$ does not shift in gate voltage.
Therefore, it is noteworthy that the points of intersection $\Gamma_\mathrm{in}=\Gamma_\mathrm{out}$ (marked by vertical lines) remain at the same position, regardless of the energy dependence introduced by $\mathcal{T}(E)$.
In both cases, the intersections shift to lower gate voltages (respectively higher energies) with increasing temperature.
This was given as explanation for the thermal shift observed by \citeauthor{brinksThermalShiftResonance2016} \cite{brinksThermalShiftResonance2016} which we also observe in our experiment (compare \autoref{fig:experimental_findings}, solid black lines).
This temperature induced shift can be seen as unaffected by the energy dependence of the tunnel barrier as the energy-dependent quantities $\mathcal{T}(E)$ and $D(E)$ simply cancel out.

\subsection{C-V spectrum of the \textit{s}1 peak}
We find for the gate-voltage dependent capacitance of charging the \textit{s}1 state of a single QD the Lorentzian
\begin{equation} 
\label{eq:theory:single_capacitance_gen}
    C(V_\mathrm{g}) = e \,\frac{\Gamma_\mathrm{in} \Gamma_\mathrm{out}}{\omega^2 + (\Gamma_\mathrm{in} + \Gamma_\mathrm{out})^2} \,\frac{\mathrm{d}}{\mathrm{d} V_\mathrm{g}} \ln \frac{\Gamma_\mathrm{in}}{\Gamma_\mathrm{out}} .
\end{equation}

\begin{figure}
    \centering
    \includegraphics[width=\columnwidth]{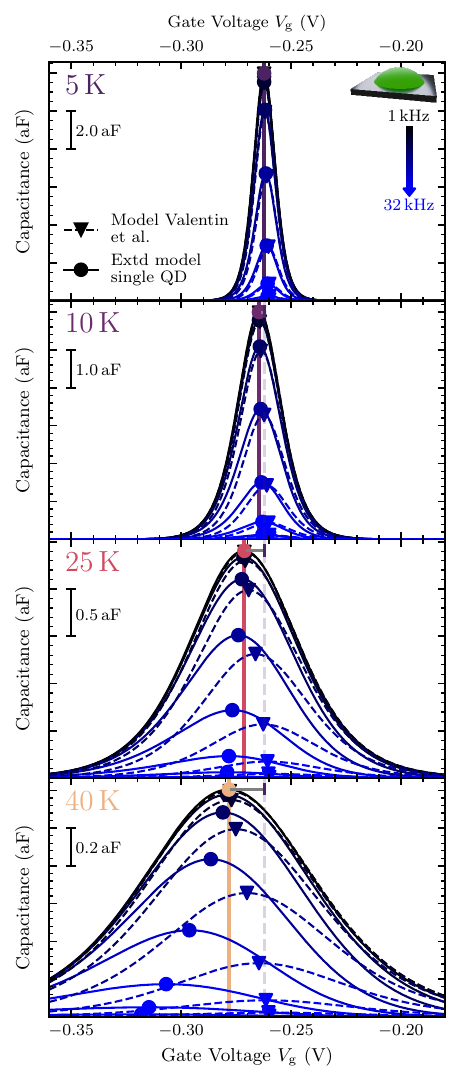}
    \caption{Simulation of the \textit{s}1 peak from a single QD for various temperatures and frequencies. The results illustrate two models: The previous one of \citeauthor{valentinIlluminationinducedNonequilibriumCharge2018} (dashed line, with peak maxima marked by triangles) and our extended model (solid line, with peak maxima marked by circles). The reversal in the direction of the frequency-induced shift at higher temperatures is only visible in the model that accounts for energy-dependent tunneling.}
    \label{fig:comparison_single_peak}
\end{figure}

The results from \autoref{eq:theory:single_capacitance_gen} allow us to simulate the C-V spectra of a single QD for the experimentally studied sample structure. We compare our model with the model of Ref. \cite{valentinIlluminationinducedNonequilibriumCharge2018}, which was similarly designed to simulate single QD C-V spectra. The comparison is shown in \autoref{fig:comparison_single_peak}, with simulation parameters chosen to match our experimental setup.

Similar to the experiment, our theoretical results show a thermal shift of the peak positions with increasing temperature at low frequencies. This behavior agrees with the explanation previously provided \cite{brinksThermalShiftResonance2016} as shown in \autoref{fig:tunnel_rates} and all low frequency experiments (\cite{brinksThermalShiftResonance2016, valentinIlluminationinducedNonequilibriumCharge2018}, this work). We also observe thermal broadening in the simulation, which seems to be more pronounced than in the experimental data. This discrepancy arises from the difference between modeling a single QD and a QD ensemble (discussed in detail in \autoref{sec:ensemble} below).

While the two models agree with each other at low frequencies, they show a distinctly different behavior at higher frequencies. At low temperatures, both models give rise to a frequency-induced shift to higher gate voltages. However, as the temperature increases, the frequency shift is opposite for the two models. In the model of Ref. \cite{valentinIlluminationinducedNonequilibriumCharge2018}, the peaks shift to higher gate voltages regardless of the temperature, whereas in our extended model the shift direction is reversed at a certain temperature and then moves to lower gate voltages as the temperature is further increased. This behavior matches the experimentally observed shifts of the peak position.

The agreement between our experimental findings and the extended model points towards an explanation of the frequency-induced shifting behavior at elevated temperatures. In the model of Ref. \cite{valentinIlluminationinducedNonequilibriumCharge2018}, the tunneling coefficient $\mathcal{T}(E)$ was assumed to be independent of energy. In contrast, our model accounts for energy dependence by incorporating an effective tunneling length calculated within the WKB approximation as described in \autoref{eq:WKB}.

\subsection{From equilibrium to nonequilibrium tunneling}

\begin{figure*}[ht]
    \centering
    \includegraphics[]{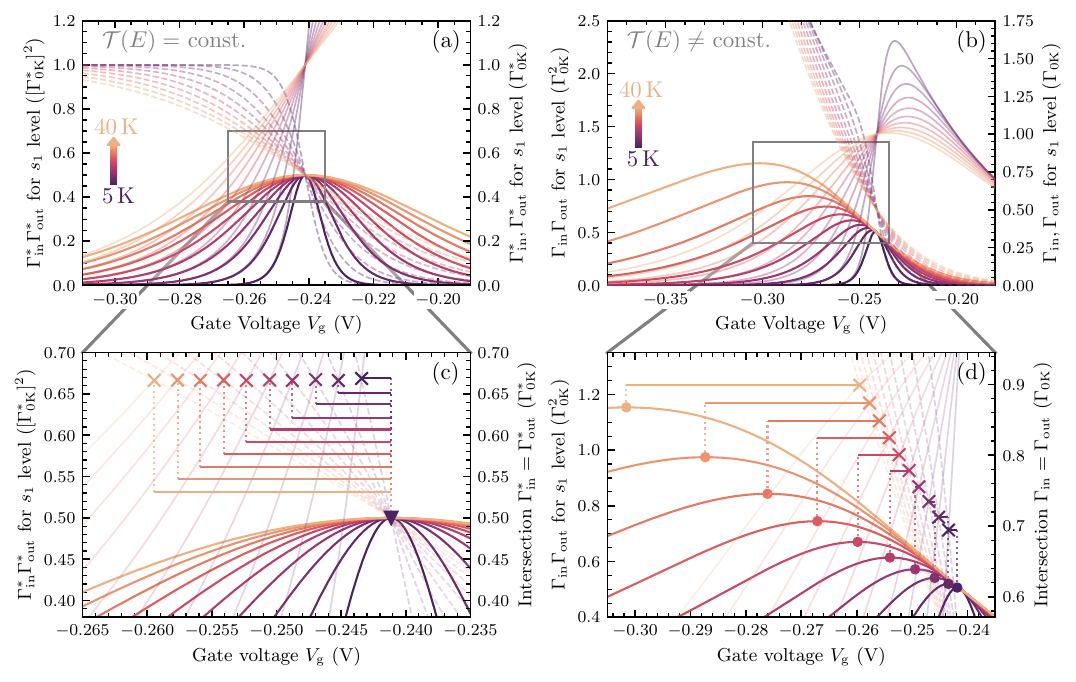}
    \caption{Product of tunneling rates (a) for energy independent tunneling and (b) including consideration of tunnel barrier's energy dependence. The rate product is plotted as a thick line and color-coded according to temperature. Using the same color scheme, the tunneling rates previously shown in \autoref{fig:tunnel_rates} are included as more transparent solid lines (for tunneling in) and dashed lines (for tunneling out).
    As zooms (c) and (d): Comparison between the maxima of the conjoint tunneling product $\Gamma_\mathrm{in} \cdot \Gamma_\mathrm{out}$ and the equilibrium positions between in and out tunneling (marked as stars at the intersection of $\Gamma_\mathrm{in}$ and $\Gamma_\mathrm{out}$). The distance between equilibrium and conjoint tunneling maximum is indicated by a colored line. In the case of energy independent tunneling (c) the shift from intersection to conjoint maximum shows only one constant direction, resembling the energy-independent tunneling model of Ref. \cite{valentinIlluminationinducedNonequilibriumCharge2018}. When considering the energy dependence induced by the tunnel barrier (d), this shift behavior changes. Now the direction reverses with rising temperature, resembling the qualitative behavior observed in the experiment. Note that the y axis are scaled different by intention to better visualize this shift.}
    \label{fig:intersections}
\end{figure*}

Having observed the difference in shifting behavior from \autoref{fig:comparison_single_peak} for low or high frequencies, we now analyze our model in these two limits.
In thermal equilibrium with the reservoir, the ratio of tunneling rates obeys detailed balance
\begin{equation} \label{eq:theory:rate_ratio}
    \frac{\Gamma_\mathrm{in}}{\Gamma_\mathrm{out}} = \frac{g_\mathrm{in}}{g_\mathrm{out}} \frac{f}{1 - f} = \frac{g_\mathrm{in}}{g_\mathrm{out}}\, \mathrm{e}^{(E_\mathrm{F}(V_\mathrm{g}) - E_{s1}) / k_\mathrm{B} T} ,
\end{equation} 
allowing us to restate \autoref{eq:theory:single_capacitance_gen} more insightful as
\begin{equation} \label{eq:theory:single_capacitance}
    C(V_\mathrm{g}) = \frac{1}{\lambda} \frac{e^2}{k_\mathrm{B} T}  \frac{\Gamma_\mathrm{in} \Gamma_\mathrm{out}}{\omega^2 + (\Gamma_\mathrm{in} + \Gamma_\mathrm{out})^2} .
\end{equation}

When the angular frequency $\omega$ is small compared to the tunneling rates, we obtain the asymptotic formula
\begin{equation} \label{eq:theory:single_capacitance_at_w_0}
    C(V_\mathrm{g}) \stackrel{\omega \rightarrow 0}{\approx} \frac{1}{\lambda} \frac{e^2}{k_\mathrm{B} T}  \frac{\Gamma_\mathrm{in} \Gamma_\mathrm{out}}{(\Gamma_\mathrm{in} + \Gamma_\mathrm{out})^2} .
\end{equation}
As the fraction of \autoref{eq:theory:single_capacitance_at_w_0} represents the squared ratio of a geometric and arithmetic mean, \autoref{eq:theory:single_capacitance_at_w_0} attains its maximum at the resonance condition $\Gamma_\mathrm{in} = \Gamma_\mathrm{out}$. 
Substituting this into \autoref{eq:theory:rate_ratio} we recover the remarkably simple expression for the peak position \cite{brinksThermalShiftResonance2016} 
\begin{equation} \label{eq:theory:single_peak_at_w_0}
    E_\mathrm{F}(V_\mathrm{peak}^{\omega \rightarrow 0}) = E_{s1} - k_\mathrm{B} T \,\ln \frac{g_\mathrm{in}}{g_\mathrm{out}}.
\end{equation}
In the limit of low measurement frequencies, our model finds the most efficient tunneling at in- and out-tunneling equilibrium. We also resemble the temperature shift given by \autoref{eq:theory:single_capacitance_at_w_0}, as observed in the experiment for low frequencies. 

In the opposite limit of large $\omega$, we find
\begin{equation} \label{eq:theory:single_capacitance_at_w_oo}
    C(V_\mathrm{g}) \stackrel{\omega \rightarrow \infty}{\approx} \frac{1}{\lambda} \frac{e^2}{k_\mathrm{B} T}  \frac{\Gamma_\mathrm{in} \Gamma_\mathrm{out}}{\omega^2} ,
\end{equation}
which attains its peak when the product $\Gamma_\mathrm{in}\Gamma_\mathrm{out}$ is maximized.
This product corresponds to the conjoint probability of successive in- and out-tunneling events, which we shall refer to as conjoint tunneling.
Differentiating $\Gamma_\mathrm{in}\Gamma_\mathrm{out}$ with respect to $E_\mathrm{F}$ results
in an implicit equation, and expanding it for small $T$ yields
\begin{equation} \label{eq:theory:single_peak_at_w_oo}
    E_\mathrm{F}(V_\mathrm{peak}^{\omega \rightarrow \infty}) \approx E_{s1} + (2 k_\mathrm{B} T)^2 \!\!\left.\frac{\mathrm{d}}{\mathrm{d}E_\mathrm{F}\!} \right|_{E_\mathrm{F} = E_{s1}} \!\!\!\!\!\!\!\!\!\!\!\!\!\!\!\!\!\!\ln (\mathcal{T} D) .
\end{equation}
Unlike \autoref{eq:theory:single_peak_at_w_0}, the shifting behavior of \autoref{eq:theory:single_peak_at_w_oo} is exclusively a result of the energy dependence of $\mathcal{T} \mathcal{D}$.
Taking the logarithm of \autoref{eq:WKB} yields a term proportional to $E_\mathrm{F}^{-1}$, resulting in a negative shift (cf. \autoref{fig:comparison_single_peak}).

How can this impact of energy dependence on tunneling behavior be understood? At low frequencies, charge fluctuations between the reservoir and the QD are largest when the rates of tunneling in and tunneling out are equal; again as already assumed \cite{brinksThermalShiftResonance2016} and calculated by the energy-independent tunneling model \cite{valentinIlluminationinducedNonequilibriumCharge2018}. When the frequency increases, we reach a regime where not all electrons manage to tunnel in and out of the QD in one ac cycle. This affects the efficiency of tunneling, and thus, the position of the measured C-V peak, differently for different $V_\mathrm{g}$.

To elaborate further on this, let us reconsider the tunnel rates. The intersections of tunnel in and out rates remain the pivotal marker for the C-V peak maximum as long as we are at low frequencies. At higher frequencies, we need to think differently about the maximum tunneling efficiency. Tunnel coupling to a QD needs to be understood as a sequential process of tunneling in and out. Thus, the maximal capacitance occurs no longer for equal tunneling in and out rates, but appears where the probability of consecutive tunneling in and out is maximized. This consecutive tunneling corresponds to the product $\Gamma_\mathrm{in} \Gamma_\mathrm{out}$ as suggested by \autoref{eq:theory:single_capacitance_at_w_oo}.

\autoref{fig:intersections} compares the tunnel rates with the conjoint tunneling product $\Gamma_\mathrm{in} \Gamma_\mathrm{out}$. It shows the results for rates that are independent of energy (a), (c) and for rates that take into account the shape of the tunnel barrier and are, thus, energy dependent (b), (d). When we compare the positions of the intersections $\Gamma_\mathrm{in} = \Gamma_\mathrm{out}$ with the maxima of conjoint tunneling at different temperatures, we observe trends that align with both experimental observations and single-dot simulations. In the case of energy-independent tunneling rates (c), the shifting behavior from the energy-independent tunneling model \cite{valentinIlluminationinducedNonequilibriumCharge2018} becomes visible: With increasing temperature, we observe a stronger growing shift of the intersections $\Gamma_\mathrm{in} = \Gamma_\mathrm{out}$ towards lower gate voltages. The maximum of the conjoint tunneling remains unchanged, regardless of the temperature.

This behavior changes if we account for energy-dependent tunneling. The shift in between intersection of in and out tunneling and the maxima of conjoint tunneling align with the peak-shift observed in the experiment or simulation considering energy dependence. At lower temperatures we observe a shift towards higher gate voltages, with increasing temperatures a change in the relative shift, and at the highest temperature a strong shift to lower gate voltages. Considering the energy dependence of the tunneling barrier proves to be the defining reason for the behavior we observe.

In summary, as the ac frequency increases, the tunneling process occurs under nonequilibrium conditions.
At low frequencies, the maximum tunnel coupling is achieved when the rates of in-tunneling and out-tunneling are equal.
However, at higher frequencies, the peaks in the C-V spectrum are rather due to a maximum of the probability for successively tunneling in and out of the QD.
This change in behavior reflects the impact of the energy-dependent tunneling on the overall tunneling dynamics.


\subsection{\label{sec:ensemble}Modeling of the inhomogeneous QD ensemble}
So far we modeled tunneling into a single QD. But our experiment features an ensemble of QDs with inhomogeneous size. This adds complexity beyond the single-QD scenario. To fully understand the experimental results, we must extend our model to account for the characteristics of the entire QD ensemble.

Our goal is to model $N$ QDs with differing size.
Assuming the QDs are not influencing each other, such that the total current is just a superposition of the single QD currents, we can simply sum their individual capacitance $C_{E_{s1}}(V_\mathrm{g})$ by \autoref{eq:theory:connect_experiment_theory}.
To consider the size variation, we assume the \textit{s}1 level energy as a Gaussian random variable $E_{s1}$ with mean $\langle E_{s1} \rangle$ and variance $\sigma^2$.
This results in an expected capacitance of
\begin{equation} \label{eq:theory:ensemble_capacity_gen}
    C_N(V_\mathrm{g}) = N \langle C_{E_{s1}}(V_\mathrm{g})\rangle ,
\end{equation}
and substituting \autoref{eq:theory:single_capacitance} into \autoref{eq:theory:ensemble_capacity_gen} yields 
\begin{equation} \label{eq:theory:ensemble_capacity}
    C_N(V_\mathrm{g}) = \frac{N}{\lambda} \frac{e^2}{k_\mathrm{B} T} \left\langle \frac{\Gamma_\mathrm{in} \Gamma_\mathrm{out}}{\omega^2 + (\Gamma_\mathrm{in} + \Gamma_\mathrm{out})^2} \right\rangle .
\end{equation}

\begin{figure}
    \centering
    \includegraphics[width=\columnwidth]{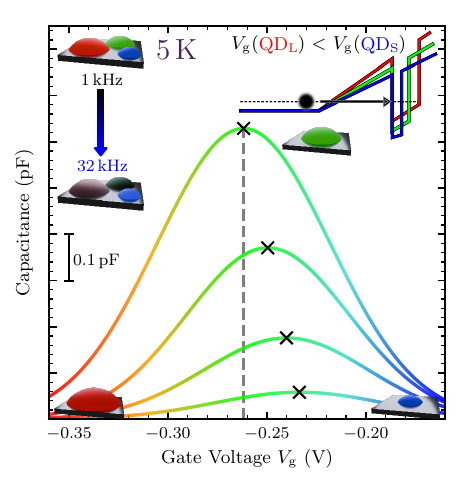}
    \caption{Effect of the energy dependent tunneling model on the frequency induced suppression of a QD ensemble. Shown are simulated C-V spectra of a QD ensemble at \SI{5}{\kelvin} for frequencies from \SIrange{1}{32}{\kilo\hertz}. The maxima (marked by a cross) shift to higher gate voltages, far stronger than visible for a single QD peak as in \autoref{fig:comparison_single_peak}. This is explained by the stronger suppression of bigger QDs, which come into resonance at lower gate voltages (compare inset). The colours of different sized QDs are picked in accordance with the size dependent energy shift of the QDs' states. 
    \label{fig:ensemble_explanation}}
\end{figure}

Using this methodology, we are able to model the C-V spectra of an ensemble of QD with inhomogeneous size. A simulation of C-V spectra at \SI{5}{\kelvin} and frequencies comparable to the experiment is shown in \autoref{fig:ensemble_explanation}. We observe a strong shift of the ensemble towards higher gate voltages (respectively lower energies). Compared to the model of a single QD, this shift is far more pronounced than in \autoref{fig:comparison_single_peak} at \SI{5}{\kelvin}.

We proceed by analyzing the peaks of \autoref{eq:theory:ensemble_capacity} for the two limiting cases. 
For small $\omega$, we first note that
\begin{equation} \label{eq:theory:logistic_distribution_omega_0}
    \frac{\Gamma_\mathrm{in} \Gamma_\mathrm{out}\,/ k_\mathrm{B} T}{\omega^2 + (\Gamma_\mathrm{in} + \Gamma_\mathrm{out})^2} \stackrel{\omega \rightarrow 0}{\approx} \ell(E_{s1} - \mu^\ast) ,
\end{equation}
where $\ell$ represents the symmetric probability density
\begin{equation}
    \ell(E) = \frac{\mathrm{e}^{E / k_\mathrm{B} T}\!/ k_\mathrm{B} T}{(1 + \mathrm{e}^{E / k_\mathrm{B} T})^2}
\end{equation}
of a zero-mean logistic distribution with scale $k_\mathrm{B} T$, and
\begin{equation} \label{eq:theory:logistic_distribution_location_omega_0}
    \mu^\ast = E_\mathrm{F} + k_\mathrm{B} T \ln \frac{g_\mathrm{in}}{g_\mathrm{out}} .
\end{equation}
Denoting the symmetric Gaussian probability density of $E_{s1} - \langle E_{s1} \rangle$ by $\phi$ yields
\begin{equation} \label{eq:theory:symmetric_intergral}
    C_N(V_\mathrm{g}) \stackrel{\omega \rightarrow 0}{\propto} \!\!\int_{-\infty}^{\infty}\!\! \mathrm{d}E_{s1}\, \phi(E_{s1} - \langle E_{s1} \rangle) \,\ell(E_{s1} - \mu^\ast) 
\end{equation}
and differentiating \autoref{eq:theory:symmetric_intergral} with respect to $E_\mathrm{F}$ effectively replaces $\ell$ by the antisymmetric derivative $\ell'$.
Since integrals over antisymmetric functions vanish, a peak is found at $\mu^\ast = \langle E_{s1} \rangle$. 
Thus by \autoref{eq:theory:logistic_distribution_location_omega_0}
\begin{equation} \label{eq:theory:ensemble_peak_at_w_0}
    E_\mathrm{F}(V_{N,\mathrm{peak}}^{\omega \rightarrow 0}) = \langle E_{s1} \rangle - k_\mathrm{B} T \ln \frac{g_\mathrm{in}}{g_\mathrm{out}} ,
\end{equation}
which generalizes the peak position from \autoref{eq:theory:single_peak_at_w_0} for symmetric distributions of QD energies.

For large $\omega$ we also find a logistic distribution, i.e.,
\begin{equation}
    \!\!\frac{\Gamma_\mathrm{in} \Gamma_\mathrm{out}\,/ k_\mathrm{B} T}{\omega^2 + (\Gamma_\mathrm{in} + \Gamma_\mathrm{out})^2} \!\stackrel{\omega \rightarrow \infty}{\approx} \frac{\mathrm{g}_\mathrm{in} g_\mathrm{out} \mathcal{T}^2 D^2 \,\ell(E_{s1} {-} E_\mathrm{F})}{\omega^2} .\!
\end{equation}
If $T$ is small, the standard deviation of $\ell$ vanishes, and $\ell$ becomes Dirac delta distribution, which reduces the expectation integral to replacing $E_{s1} = E_\mathrm{F}$, i.e.,
\begin{equation}
    C_N(V_\mathrm{g}) \underset{T\rightarrow 0}{\stackrel{\omega \rightarrow \infty}{\propto}} \left. (\mathcal{T} D)^2 \right|_{E_{s1} = E_\mathrm{F}} \!\mathrm{e}^{-\frac{1}{2} \frac{(E_\mathrm{F} - \langle E_{s1} \rangle)^2}{\sigma^2}} .
\end{equation}
Again, this yields an implicit equation for the peak position, which is approximated for small $\sigma$ and $T$ by
\begin{equation} \label{eq:theory:ensemble_peak_at_w_oo}
    E_\mathrm{F}(V_{N,\mathrm{peak}}^{\omega \rightarrow \infty}) \approx \langle E_{s1} \rangle + 2 \sigma^2 \!\!\left.\frac{\mathrm{d}}{\mathrm{d}E_\mathrm{F}\!} \right|_{E_\mathrm{F} = \langle E_{s1} \rangle} \!\!\!\!\!\!\!\!\!\!\!\!\!\!\!\!\!\!\!\!\!\!\ln ([\mathcal{T} D]_{E_{s1} = E_\mathrm{F}}) .
\end{equation}
Thus, the ensemble introduces an additional energy dependent shift similar to \autoref{eq:theory:single_peak_at_w_oo}.
However, as $E_{s1}$ is replaced by $E_\mathrm{F}$ prior to differentiation, the logarithm in \autoref{eq:theory:ensemble_peak_at_w_oo} becomes proportional to $-\sqrt{-E_\mathrm{F}}$ instead of $E_\mathrm{F}^{-1}$, resulting in an opposite shift (compare \autoref{fig:comparison_single_peak} and \autoref{fig:comparison_theory_experiment}).

Again, an explanation of this additional ensemble effect can be given by considering an energy dependence in tunneling. As illustrated in the inset of \autoref{fig:ensemble_explanation}, the size of a QD impacts the necessary band tilt (applied gate voltage) where the QD energy level comes into resonance with the Fermi-level of the reservoir. Bigger, red-shifted dots need a lower gate voltage than smaller, blue-shifted dots. This impacts the tunnel barrier carriers need to surpass. Carriers tunneling into a state of bigger dots face an effectively larger barrier than carriers tunneling into smaller dots. If the frequency is now increased and less carriers participate in tunneling, the QDs with a larger tunnel barrier get suppressed more strongly. Consequently, increasing the frequency shifts the position of the C-V peak to higher gate voltages, where primarily the smaller, less suppressed QDs contribute.

\subsection{Comparison between extended model and experiment}

\begin{figure}
    \centering
    \includegraphics[width=\columnwidth]{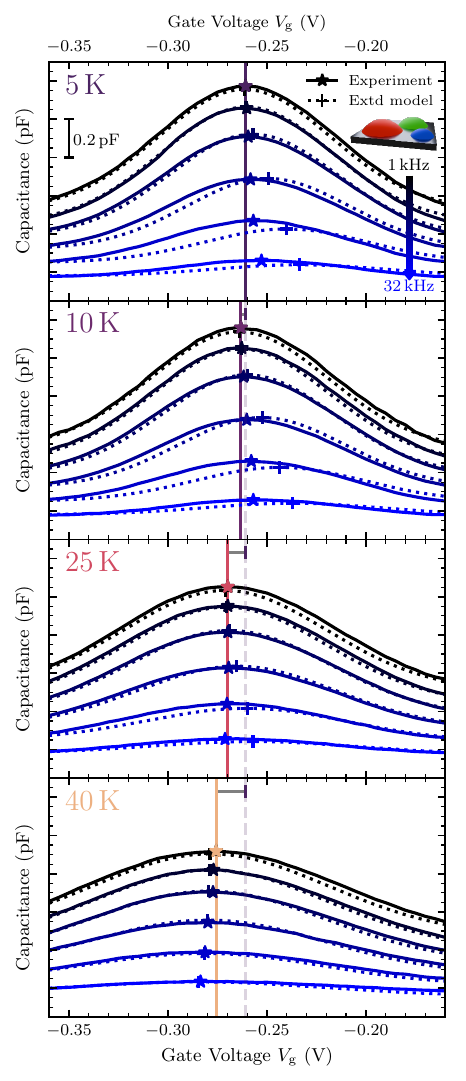}
    \caption{Comparison between experimental C-V spectra (straight lines with stars marking the peak maxima) and the modeled spectra considering energy dependent tunneling in a QD ensemble (dotted line, crosses). Only the data of the  first electron charging peak is compared; the underlying baseline and tail of the second peak is fitted and subtracted out of the experimental C-V spectra. The model is able to resemble the experiment well, and agrees with the direction change of the frequency induced peak-shift at higher bath temperatures.}
    \label{fig:comparison_theory_experiment}
\end{figure}

With all these effects incorporated into our extended model, we compare our theoretical results with the experimental data. The parameters for the sample geometry and QD ensemble were selected as follows: $d_0=\text{\SI{35}{\nano\meter}}$, $d_\mathrm{tot}=\text{\SI{268}{\nano\meter}}$ were selected from the geometrical sample structure. The built-in voltage $V_\mathrm{bin}$ was chosen to coincide with the band gap of GaAs. Remaining parameters were fitted to match the experimental data. \autoref{fig:comparison_theory_experiment}  presents a comparison between the experimental data and the simulation results obtained using our comprehensive model. Peak positions are marked in both the experimental data and the simulated results for easy comparison.

The width of the ensemble peak and the qualitative direction of the frequency-induced shift are very similar in the model and the experiment. At lower temperatures, the model predicts a stronger shift than what was observed experimentally. However, at higher temperatures, the model and experimental results align well.

The qualitative agreement demonstrated by our extended model underscores its robustness in capturing the essential physics of the system. Nevertheless, quantitative deviations in the frequency-induced shifting behavior suggest opportunities for further refinement. One potential factor is the precise determination of the distance $d_0$ between the charge reservoir and the QD, which plays a crucial role in the calculation of tunneling rates via \autoref{eq:theory:tunneling_rate_in} and \autoref{eq:theory:tunneling_rate_out}. In our model, this distance is referenced to a distinct band edge, as depicted in \autoref{fig:sketch_bandstructure}. However, in an actual experimental setting, the band edge exhibits bending, implying that $d_0$ may vary slightly. The random character of dopent distribution is an uncertain factor adding to this. A more detailed consideration of this effect could further enhance the quantitative accuracy of our model.

Another promising direction for refining the model is the inclusion of strain effects, which influence the energy dependence of the system but have not yet been incorporated. Strain modifies the band structure by introducing curvature, thereby altering the energy response to variations in gate voltage. This effect originates from the lattice mismatch between materials and is particularly relevant in the vicinity of QDs, where it drives the self-organized growth of InAs QDs on a GaAs matrix. Given the inherent size dispersion of QDs, it is reasonable to assume that strain effects vary across the ensemble. Larger QDs, which typically contain a higher indium fraction, are expected to experience increased strain, leading to more pronounced band bending. This curvature effectively modifies the tunneling barrier, potentially reducing its width for In-rich QDs. Consequently, these larger QDs, located on the lower gate voltage side of the ensemble, would experience weaker suppression. By incorporating this effect, the model could achieve an even more precise representation of the experimental observations, particularly in capturing the shift toward higher gate voltages.


\section{\label{sec:conclusion}Conclusion}

Our study presents findings in the C-V spectroscopy of reservoir-coupled QDs, uncovering a previously unreported frequency-induced shifting behavior in the single electron peak. These experimental observations are explained by an extended model based on a master equation approach that captures the complex interplay between temperature, frequency, and QD ensemble effects. With our model we are able to qualitatively resemble the experimental behavior. On the base of our model we gained insights in the tunnel coupling between reservoir and QD.

In comparison to the energy-independent tunneling model \cite{valentinIlluminationinducedNonequilibriumCharge2018}, the relevance of considering energy dependence in tunnel coupling became clear. We introduced a distinction between a balance of tunneling in and out for low frequencies and conjoint tunneling at high ac frequencies as explanation for the impact of energy dependence. The extension of our model to treat an inhomogeneous ensemble of QDs revealed another energy-dependent tunneling effect: due to differences in the effective height of the tunnel barrier for differently sized QDs, larger QDs experience stronger suppression as the frequency increases. This ensemble effect dominates the frequency-dependent response at lower temperatures but becomes less significant as the temperature increases.

Our results highlight the importance of accounting for the energy dependence of the tunnel barrier, especially in non-equilibrium scenarios such as those induced by high frequencies. This energy dependence is crucial for accurately describing the behavior of QD ensembles and their response to frequency and temperature variations. Future work could explore how variations in the barrier’s shape impact the system's response, potentially enabling tunable temperature or frequency-induced shifts in QD states during C-V measurements.

By implementing structural modifications, it may be possible to tailor these shifts effectively, as demonstrated by Ref. \cite{korschElectronTunnelingDynamics2020}. Such adjustments could lead to the development of resonant tunnel structures with non-monotonic energy dependencies.
These are of interest for applications like energy harvesting using QDs \cite{sothmannThermoelectricEnergyHarvesting2014, thierschmannThermoelectricsCoulombcoupledQuantum2016, jordanPowerfulEfficientEnergy2013}. Additionally, pursuing designs for charge-stable structures at elevated temperatures could open up possibilities for device innovation and performance enhancement of QD based classical and quantum devices.

Achieving a quantitative reproduction of C-V measurements remains an ambitious challenge. The interplay of a variety of influence factors, like segregation driven doping profiles or strain induced band bending, underlines the complexity of this task. While the present model already provides substantial insight and strong qualitative agreement, further refinements offer exciting avenues for improvement. These refinements would not only enhance the quantitative alignment of the model but also deepen our fundamental understanding of the underlying processes. 

\acknowledgments
We acknowledge funding by BMBF QR. N project 16KIS2200, QUANTERA BMBF EQSOTIC project 16KIS2061, as well as DFG excellence cluster ML4Q project EXC 2004/1 and funding by the Deutsche Forschungsgemeinschaft (DFG, German Research Foundation) Project No. 505408069. The authors would also like to acknowledge the DFH/UFA CDFA-05-06 Nice-Bochum Research School.
BS acknowledges funding by the Deutsche Forschungsgemeinschaft (DFG, German Research Foundation) Project No. 278162697-SFB 1242.

\bibliography{library.bib}

\end{document}